\documentclass[12pt]{iopart}

\usepackage{graphicx}

\newcommand{\be}{\begin{equation}}
\newcommand{\ee}{\end{equation}}

\newcommand{\vecp}{{\mathbf p}}
\newcommand{\vecq}{{\mathbf q}}

\newcommand{\x}{{\mathbf x}}

\newcommand{\J}{{\mathbf J}}

\newcommand{\vQ}{{\mathbf Q}}

\newcommand{\W}{{W}}
\newcommand{\A}{{\hat{A}}}

\newcommand{\opA}{{\hat{A}}}

\newcommand{\oprho}{{\hat{\rho}}}

\newcommand{\M}{{\mathbf M}}

\begin{document}
\title{Negativity witness for the quantum ergodic conjecture}
\author {Alfredo M. Ozorio de Almeida\footnote{ozorio@cbpf.br}}
\address{Centro Brasileiro de Pesquisas Fisicas,
Rua Xavier Sigaud 150, 22290-180, Rio de Janeiro, R.J., Brazil}

\begin{abstract}
25/02/2014

For an ergodic system, the time average of a classical observable
coincides with that obtained via the Liouville probability density,
a $\delta$-function on the energy shell. Reinterpreting this distribution as a Wigner function,
that is, the Weyl representation of a density operator, the quantum ergodic conjecture
identifies this classical construction as an approximate representation of the quantum eigenstate
of the same energy. It is found that this reasonable hypothesis, as far as expectations for observables
are concerned, does not satisfy the requirement of positivity, so that the $\delta$-function
on the shell cannot be a true Wigner function. This result was first presented by N. Balazs
for the special case of a single degree of freedom, such that the system is simultaneously
integrable and ergodic. Here, it is shown, for a $\delta$-function on an arbitrary curved energy shell
in a $2N$-dimensional phase space, that there exists a positive operator 
for which the conjectured Wigner function predicts a negative expectation.

\end{abstract}

\maketitle
\section{Introduction}

Representations of quantum mechanics in phase space are specially apt 
to highlight both correspondences and dissonances with classical mechanics.
Among these, the Wigner function \cite{Wigner}, 
\be
W(\x)=\frac{\rho(\x)}{(2\pi\hbar)^N}=
\frac{1}{(2\pi\hbar)^N} \int d{\vQ}~ \langle \vecq+\frac{1}{2}\vQ |\oprho| \vecq-\frac{1}{2}\vQ\rangle~
e^{-i\vQ\cdot \vecp/ \hbar},
\ee
has played a major role. It can be considered as the Weyl symbol
of the density operator, $\hat \rho$, on a {\it classical phase space}, 
$\mathbf{R}^{2N}: \x=(\vecp,\vecq)$. Notwithstanding its express purpose of
calculating the expectation of arbitrary operators, $\hat A$, from their
Weyl symbol, $\A(\x)$, in a classical manner, i.e.,
\begin{equation}
\langle \opA \rangle = \int d\x \;W(\x)\;A(\x) \ ,
\label{expectation}
\end{equation}
the true quantum nature of the Wigner function entails the existence of regions where it is negative.
Indeed, according to Hudson's theorem \cite{Hudson}, the only instances
of pure states with a non-negative Wigner function are the
Gaussians that represent coherent states. The existence of such negative regions 
in no way affects the {\it positivity} of the density operator itself: 
This is the guarantee of a positive expectation $\langle \opA \rangle$,
if $\opA$ only has positive eigenvalues. 

Even so, it is by no means evident whether an operator $\oprho_{\pi}$ that is 
defined by the inverse Weyl association from a putative Wigner function, $W_{\pi}(\x)$, 
will indeed satisfy this essential requisite. For instance, if the evolution of a 
known bona fide Wigner function is calculated within an approximate scheme,
can one be sure that no negative eigenvalues of the original density operator
have become negative? It is not at all clear how to guarantee positivity.
Instead, it may be more practicable 
to falsify this condition. The idea is to employ a known density operator, $\oprho_{\omega}$,
and hence positive, with its Wigner function, $W_w(\x)$, as a {\it negativity witness}. 
That is, by verifying that the expectation of the operator $\oprho_{\pi}$,
\begin{equation}
\int d\x \;W_w(\x)\;W_{\pi}(\x) = \langle \oprho_{\pi}\rangle_w  < 0 \ ,
\label{negative}
\end{equation}
one concludes that the operator $\oprho_{\pi}$ is not positive and $W_{\pi}(\x)$ is not a true 
Wigner function.

A celebrated instance of such a putative Wigner function results from the 
{\it quantum ergodic conjecture} (QEC) of Voros \cite{Voros76} and Berry \cite{Berry77b}: 
The proposed Wigner function is simply identified with the classical {\it Liouville probability density}
for the $E$-energy shell of the Hamiltonian with Weyl symbol $H(\x)$
\footnote {$H(\x)$ coincides with the classical Hamiltonian, $H_C(\x)$, in the important case where
$H(\x)=p^2/2m + V(q)$. For general mechanical observables, $\opA$, the approximation for the Weyl symbol, $A_C(\x)\approx A(\x)$, only holds asymptotically in the semiclassical limit.},
\be
W_{\pi}(\x) = \frac{\delta(E-H(\x))}{\int d\x ~\delta(E - H(\x))},
\label{ergodic}
\ee
where $\delta$ denotes the Dirac delta-function.
The assumption is that the classical Hamiltonian is {\it ergodic}, so that the time average of any
classical observable, $A_C(\x)$, evolved by $H(\x)= H(p,q)$ according to Hamilton's equations,
coincides with the classical phase space average \eref{expectation} for the classical probability density \eref{ergodic}. The conjecture is then that for most eigenenergies
this same ergodic distribution, $W_{\pi}(\x)$, can be reinterpreted as an approximate quantum 
Wigner fuction that represents the $E$-eigenstate of the corresponding quantum Hamiltonian.

It is a consequence of this conjecture that the quantum expectation values 
for arbitrary quantum observables $\opA$, with the Weyl symbol, 
$A(\x)\approx A_C(\x)$, then coincide with the ergodic classical averages.
This is, indeed, confirmed by Shnirelman's theorem 
\cite{Shnirelman, Verdiere, Zelditch} for a majority of eigenstates, 
albeit with severe restrictions for the allowed classes both of the observables and of the ergodic systems. It is important to note that, while the Shnirelman theorem concerns individual states 
and expectation values, it is not formulated directly in terms of Wigner functions.

One could question the value of aiming a conjecture at a full representation
of a quantum ergodic state, such as the Wigner function, 
if all it did were to determine good expectation values. 
However, QEC is stronger than Shnirelman's theorem, with consequences 
that are much more subtle than the gross
correspondence of expectations to their classical values. 
Most notable, so far, is the deduction by Berry \cite{Berry77b} 
of local statistical correlations of wave functions, that are directly applied 
to the interpretation of experiments in quantum dots (see \cite{Jalabert} and references therein)
Eventually, the ever increasing experimental refinement in manipulating quantum states
may allow one to access other delicate properties, such as interference phenomena,
for eigenstates of classically ergodic Hamiltonians.

Let us enumerate some a priori restrictions that disallow 
the putative Wigner function for the QEC:
First, there is the mentioned theorem by Hudson \cite{Hudson}, that is,
one can define the $\delta$-function in \eref{ergodic} as a limit of positive functions,
so that $W_{\pi}(\x)$ becomes a positive function in phase space, but it is not a Gaussian. Furthermore, there is the fact that correlations 
of a true Wigner function for an individual pure state are invariant 
with respect to Fourier transformation \cite{Chountasis, OVS}. 
Thus, a large scale structure (the energy shell itself) must be accompanied 
by oscillations with large wave vectors,
that is, at very fine scales. These fine oscilations may be associated to the
{\it sub-Planck} structures of the Wigner function discussed by Zurek \cite{Zurek}.
On the other hand, Wigner functions that are peaked on the classical energy shell
have been derived semiclassically, but for mixed states within a narrow energy window.
Even so, the appropriate form is given in terms of highly oscillatory Airy functions, 
rather than $\delta$-functions \cite{Berry89,Report,TosLew}.
Further Wigner oscillations arise if one reduces the width of the energy window, 
so that one may presume an individual pure Wigner function to have a complex detailed pattern,
full of negative regions and interferences, far removed from the QEC. It follows that expectation values can only coincide with those of classical mechanics
as a result of the smoothing effect, on the fine Wigner oscillations 
within the average (\ref{expectation}), of the slowly varying
functions corresponding to typical mechanical observables. 

The most serious setback is the suspicion that positivity is violated, 
which does not even allow $\W_\pi(\x)$ to be a mixed state. 
For one degree of freedom, Nandor Balazs employed a positivity witness,
just as in \eref{negative}, to prove that the operator $\oprho_\pi$ is not positive. 
However, this result, which concerns a $\delta$-function along a curve,
remains largely ignored in an appendix of \cite{Balazs}.
One should note that a classical Hamiltonian for a single degree of freedom
is only ergodic in a trivial sense, being as it is also integrable.
Ergodicity only becomes a characteristic of chaos for $N \geq 2$
and it has remained an issue whether some generalization of Balazs's positivity witness
will evince negativity for the Wigner function \eref{ergodic}, that is
concentrated on a higher dimensional codimension-1 manifold, for general $\mathbf{R}^{2N}$.

The alternative that is here followed is to replace Balazs's witness,
based on a superposition of position states, by a negative {\it Schr\"odinger cat state},
i.e a superposition of apropriate coherent states. This allows for the pinpointing of local
structure of the shell, whatever the dimension, and hence a general detection
of negativity. The new witness is presented in section 2, followed by a rederivation
of Balazs's result. Intuitively, one may simplify the energy shell by its local tangent
within each of the three regions of a cat state where the Wigner function has an appreciable amplitude.
This is confirmed by a full inclusion of curvature in the Appendix.
Then it is shown in section 3 that a local separation of phase space
coordinates is allowed by symplectic invariance, which reduces the evaluation 
of the multidimensional negativity integral  \eref{negative}
to the case of a single freedom. Implications and possible resolutions of
the dilemmas concerning the quantum ergodic conjecture are discussed in section 4.

\section{A local witness}

Let us consider the expansion of the Weyl symbol for the Hamiltonian about an arbitrary point, $\x_0$,
of the energy shell, $H(\x)=E$. Shifting the origin to $\x_0$ leaves the Hamiltonian invariant, i.e.
$H'(\x'=\x-\x_0)=H(\x)$, so that, expanding the new Hamiltonian around the origin 
(while dropping the primes) we have,
\be
H(\x)=H(0) + \nabla H \cdot \x + \frac{1}{2} ~\x \cdot \mathbf{H} ~\x + ...,
\label{expansion} 
\ee
where $\nabla H$ is the gradient of the Weyl Hamiltonian at the origin and $\mathbf{H}$
is the Hessian matrix of second derivatives at this point. The Weyl Hamiltonian is also
invariant with respect to the group of {\it metaplectic transformations}
\cite{Bargmann, KramMoshSel, GuilStern, Voros76, Voros77, Littlejohn86, deGosson06},
that is, it transforms according to the corresponding classical {\it symplectic transformation}:
$\x \mapsto \x' = \M~\x$, where $\M$ is a {\it symplectic matrix} \cite{Arnold}.
There always exists a symplectic transformation which identifies the direction of the
gradient vector with the new $p_1$-axis, or simply the $p$-axis in the simple case of $\mathbf{R}^2$,
which will be assumed henceforth in this section. 

The tangent to the energy shell is then in the $q$-direction, 
so that again droping the primes for the new coordinates and the new Hamiltonian,
the energy shell in the quadratic expansion \eref{expansion} may be put in the form
\be
H(\x)-E = -\alpha~p + (a~p^2 + b~q^2 + c~pq) = 0.
\ee 
Having chosen the expansion point on the (large) energy shell so that the gradient ($\nabla H=-\alpha$) is not small,
one has $\alpha>>ap+cq$, leading to the parabolic approximation for the energy shell:
$p^2 \approx (b/ \alpha)q^2$. Locally, this is equivalent to a more convenient circular approximation,
in terms of the radius of curvature, $R=\alpha/\omega$,
\be
H(\x) = \frac{\omega}{2}[(p-R)^2 + q^2]
\label{circle}
\ee
so that $E=\omega R^2$.
\footnote{The invariance of the Hamiltonian with respect to canonical transformations
allows for the multiplication of $p$ together with the division of $q$ by the same dimensional
factor, without altering the value of classical actions. Thus, it will be assumed here 
that $p$ and $q$ have the same dimension, coinciding with that of $\hbar^{1/2}$.}

The negativity witness is now chosen to be a superposition of a pair of coherent states 
with opposite phases, vulgarly known as an an odd {\it Schr\"odinger cat state}, 
with the Wigner function
\begin{equation}
 W_{w}(\x) = 
\frac{e^{-p^2 / \hbar }}{2\pi \hbar \,(1 - e^{-Q^2/\hbar})}
\left[e^{-(q-Q )^2 / \hbar } +
      e^{-(q+Q )^2 / \hbar } - 
    2 e^{-q^2/ \hbar } \cos \frac{2}{\hbar} Qp \right].
\label{Wigcat}
\end{equation}
Its relation to the locally circular energy shell is displayed in Fig.1, where the
intensity increases with darkness (irrespective of the sign), while the nodes are white. 
This choice of witness is made because antisymmetry selects the origin 
as the point of maximum intensity and the negative sign
accompanies the $q$-axis until the positive Gaussians centred at $\pm Q$ are approached.
The latter have a width of $O(\hbar^{1/2})$ and it will be assumed that $Q$ is of $O(\hbar^\gamma)$,
where $0 < \gamma < 1/2$, so that the average \eref{negative} depends on a classically small
neighbourhood of the energy shell.

\begin{figure}[htb!]
\centering
\includegraphics[height=9cm]{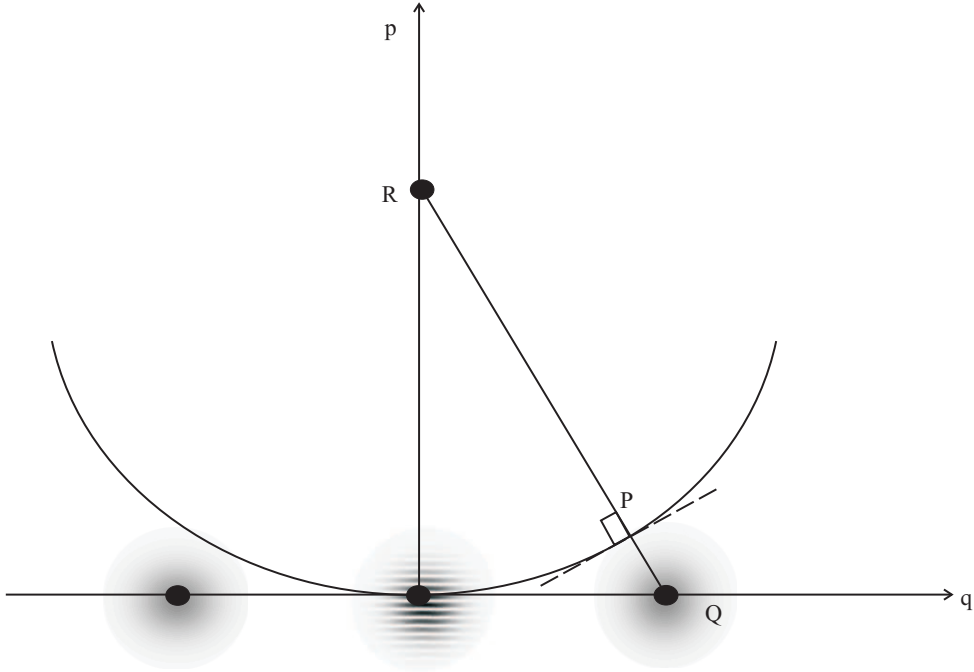}
\caption{Wigner function for negativity witness, optimally placed on a locally circular energy shell.
The tangent to the shell at the origin coincides with the trough of maximum negativity
in the central oscilatory region, whereas the circle avoids the positive maxima at the centres of the coherent states.}
\label{Fig1}
\end{figure}

Now it is convenient to separate the intrinsically positive contribution to the average \eref{negative}
due to the pair of individual coherent states, i.e.
\be
\langle {\oprho_\pi}\rangle^+ \equiv C \int dpdq~ \delta\left[ \frac{\omega}{2}((p-R)^2 + q^2 - R^2) \right]
e^{-(p^2 + (q-Q)^2)/ \hbar} ,
\label{posav}
\ee
from the contribution of the interference term,
\be
\langle {\oprho_\pi}\rangle^- \equiv C \Re \int dpdq~ 
\delta\left[ \frac{\omega}{2}((p-R)^2 + q^2 - R^2) \right]
e^{-(p^2 + q^2+2iQp)/ \hbar} ,
\label{negav}
\ee
such that $\langle {\oprho_\pi}\rangle = \langle {\oprho_\pi}\rangle^+ - \langle {\oprho_\pi}\rangle^-$.
In order to evaluate \eref{negav} one notes that, because of the Gaussian term centred on the origin, the circle can be approximated by its tangent, the $q$-axis within the $\delta$-function in \eref{negav}, which reduces to $\delta(\omega Rp)$ and hence
\be
\langle {\oprho_\pi}\rangle^- \approx \frac{C}{\omega R} \int dq~ e^{-q^2/ \hbar } = 
\frac{C}{\omega R} \sqrt{\hbar\pi}.
\label{negav2}
\ee

To evaluate \eref{posav}, one takes the origin to the point $Q$ on the $q$-axis 
and again performs a symplectic rotation of the coordinates. Then the line $\overline{RQ}$
becomes the new $p$-axis, whereas the new $q$-axis becomes the tangent of the circular energy shell
at the intersection with $\overline{RQ}$. This is the dashed line in Fig.1.
Identifying the distance $\overline{QR}$ with the value of the new $p$-coordinate
at the intersection, then specifies the distance between the point $Q$ and the intersection as
\be
P \equiv \sqrt{R^2 + Q^2} - R \approx \frac{Q^2}{2R};
\label{intersection}
\ee
where it is recalled that $R$ is a classical variable while $Q=O(\hbar)$.
It is again justified to restrict the circle by its tangent, now at the intersection point $(P,0)$,
so that \eref{posav} becomes 
\be
\langle {\oprho_\pi}\rangle^+ \approx C\int dpdq~ \frac{\delta(p-P)}{2\omega \sqrt{R^2+Q^2}} ~ e^{-(p^2+q^2)/ \hbar }
= \frac{C\sqrt{\pi\hbar}}{\omega \sqrt{R^2+Q^2}}~ e^{-P^2/ \hbar}.
\label{posav2}
\ee
Thus, the above approximation for $P$, together with $Q= K\hbar^{1/5}$ leads to
\be
\langle {\oprho_\pi}\rangle^+ \approx \frac{C\sqrt{\pi\hbar}}{\omega R}~ 
\exp \left(- \frac{K^4}{4R^2\hbar^{1/5}}\right).
\label{posav3}
\ee

The conclusion is that $\langle {\oprho_\pi}\rangle^+$ is smaller than $\langle {\oprho_\pi}\rangle^-$
by a factor that grows  exponentially with $\hbar$. Therefore, $\langle \oprho_\pi\rangle <0$
and hence ${\oprho_\pi}$ cannot represent a true density operator. So there is no doubt that the
exponential factor betweeen the positive and the negative term is immune to curvature corrections,
a full calculation for the circle in the Appendix confirms the result of this section.

The negativity witness employed by Balazs \cite{Balazs} also utilizes a negative region
of the Wigner function due to interference, but for a pair of position states, $|q_1\rangle-|q_2\rangle$,
rather than coherent states. Thus, the corresponding positive though unbounded operator is represented by
\be
W_B(\x) = \delta(q-q_1) + \delta(q-q_2) - \delta\left(q-\frac{q_1+q_2}{2}\right) \cos\frac{(q_2-q_1)p}{\hbar}.
\ee 
The positions are then chosen so that the point $\left(0, \frac{q_1+q_2}{2}\right)$ lies on the energy shell
with a vertical tangent, i.e with $p=0$. It is then shown that the negative contribution at this 
tangency is larger than the pair of intersections of $q=q_1$ with the shell.

Both choices of witness depend on the curvature of the energy shell for the detection of negativity
in the putative quantum ergodic Wigner function \eref{ergodic}. Indeed, a $\delta$-function on any
straight line is the Weyl symbol for an unbounded but positive operator just as $|q\rangle\langle q|$,
so that its average for any witness is positive.

\section{Higher dimensions}

The negativity witness employed by Balazs in \cite{Balazs} is not easily generalized 
to higher dimensional phase spaces. The Wigner function for a position state $|q_1\rangle$
is again $\delta(q-q_1)$, but this is now a $\delta$-function on an $N$-dimensional
plane, while the interference is also located on the plane $q=(q_1+q_2)/2$. 
The problem is that the $q_1$-plane intersects the energy shell 
along $(N-1)$-dimensional manifolds, instead of at isolated points.
For this reason, it is simpler to generalize the new witness presented in section 2,
which averages over the purely local structure of the energy shell, whatever the dimension.

Let us again consider the expansion \eref{expansion} of the Hamiltonian about the
origin, placed on the energy shell, but now for a general phase space, $\mathbf{R}^{2N}$.
A symplectic transformation is then chosen that, not only identifies the $p_1$-axis with
$\nabla H$, but also selects the $q_1$-direction as that of the trajectory at the origin: 
$\dot{\x}=\J\nabla H$, that is, according to Hamilton's equations. Then, repeating
the approximations that led to \eref{circle} within the $(p_1, q_1)$ plane, leads to
\be
H(\x) = \frac{\omega}{2}[(p_1-R)^2 + {q_1}^2] + \frac{1}{2} \x \cdot {\mathbf H}'~\x + ...~.
\label{circle2n}
\ee
${\mathbf H}'$ takes care of the remaining terms of the second order expansion at the origin,
so that each term is at least linear in some component of $\x'=(\x_2,...,\x_n)$.
It follows that the 2-dimensional plane $\x'=0$ is invariant to this order of approximation.
\footnote{If one chooses $\x_0$ on a periodic orbit, which generally belongs to a family,
parametrized by the energy, the local invariant plane for \eref{circle2n} will be tangent
to the 2-dimensional ring formed by the family of orbits (see e.g. \cite{OzLivro})}
Thus, the gradient, $\nabla H(\x)$, at all the points $\x= (\x_1, \x'=0)$, also lies
in this plane, so that the tangent plane will be the Cartesian product of $\J \nabla H(\x)$ with $\x'$. 

The next step is then to choose as positivity witness the tensor product of the 2-dimensional witness \eref{Wigcat}
with the coherent state centred on the origin for the remaining variables, $\x'$. 
Thus, the Wigner function for the witness is the product of \eref{Wigcat}, 
in the variables $(p_1, q_1)$, with
\be
{W'}_w(\x')=\frac{1}{(\pi\hbar)^{N-1}} \prod_{n'=2}^n \exp\left[-\frac{1}{\hbar}({p_{n'}}^2+{q_{n'}}^2)\right].
\ee
The product of this $(2n-2)$-dimensional Wigner function with each of the separate coherent states
in \eref{Wigcat} is just the positive definite Wigner function for a $2n$-dimensional coherent state.
Hence, the resulting $2n$-dimensional Schr\"odinger cat Wigner function is positive along the
tangent plane to the energy shell at $P$, which is the Cartesian product of the tangent shown in Fig.1
with the $(2n-2)$-dimensional plane, $\x'=0$. Even so, just as before, the resulting positive contribution 
to the average of $W_{\pi}(\x)$ is counterbalanced by the integral over the negative interference term
along the tangent plane at the origin, namely $p_1=0$. Indeed, the ratio between the moduli 
of the negative term $\langle {\oprho_\pi}\rangle^-$ and the positive term $\langle {\oprho_\pi}\rangle^+$ is dominated by the same exponential factor in \eref{posav3}.

\section{Discussion}

Positivity is very hard to acertain directly, but one may be fairly confident that
standard semiclassical approximations are safe. For an eigenstate of 
the harmonic oscillator, the semiclassical approximation is very close to the exact
Wigner function \cite{Berry77} and one can always picture the semiclassical evolution
as a recreation of the Wigner function from the geometry of the corresponding classical manifolds.
The semiclassical extention to nonunitary evolution of open systems requires more care.
Even so, in the case of Markovian evolution, the coarse-graining of the Wigner function
takes an original pure state into a mixture \cite{AlmRiBro, BroAlm10, AlmBro11}. 
This is shown to amount to a smoothing of the Wigner function 
that is quite similar to a local average of initial eigenstates, which would be safe, 
if each of the individual pure states is positive.

The dilemma regarding the QEC is highlighted by the contrast with the semiclassical theory
for integrable systems. If the correct structure of the Wigner function for an eigenstate
were not well established \cite{Berry77, OAH1}, it would be reasonable to conjecture it
as a $\delta$-function on the corresponding classical torus. This would supply 
the correct expectations and Berry \cite{Berry77b} showed further that local correlations
for wave functions are also obtained. However, in the case of a single degree of freedom,
this proposed Wigner function would coincide with the QEC, which does not pass the negativity test.
In this case, one can immediately compare the crude conjectured Wigner function with the rich
structure of the full semiclassical approximation. There is no doubt that chaotic eigenstates
in higher dimensions also have interferences and oscilations, which on the whole guarantee
the property of positivity, but which have proved to be a lasting and formidable theoretical challenge.

The fact, that the wave function correlations deduced by Berry \cite{Berry77b} from the QEC
have become an important instrument in the mesoscopic theory of quantum dots \cite{Jalabert},
emphasizes the awkwardness of the present theoretical predicament: The interpretation of 
experimental results is based on a Wigner function that is not positive, as well as suffering from
further drawbacks discussed in the introduction. In view of the persistent difficulty in obtaining 
a satisfactory and more complete description of the Wigner function of a classically chaotic eigenstate,
an alternative may be to recast the quantum ergodic conjecture in  some other representation.
The requirement is that positivity is not violated, while the desired quantum features, such
as wave function correlations, are still described in a satisfactory way.

\appendix

\section{Curvature corrections}

The approximation of a circular energy shell by its tangent tends to increase the
negative average $\langle {\oprho_\pi}\rangle^-$ given by \eref{negav}, 
because the tangent lies entirely within a negative region of the 
witness Wigner function, whereas the circle eventually passes through
positive regions. To show that the effect of curvature is truly negligible 
in the semiclassical limit, the full calculation for a circle is presented here.
Without the approximation by the tangent to the energy shell at the point $P$ of Fig 1,
we may rewrite \eref{posav} as
\be
\langle {\oprho_\pi}\rangle^+ = C \int dpdq~ \delta\left[ \frac{\omega}{2}(p^2 + q^2 - R^2) \right]
e^{-[(p+\sqrt{R^2+q^2})^2 + q^2)]/ \hbar} .
\label{posav4}
\ee
Then, passing to circular coordinates, such that $p^2 + q^2 = r^2$,
this becomes
\begin{eqnarray}
\langle {\oprho_\pi}\rangle^+ &= \frac{C}{2} ~e^{-(R^2 + Q^2)/ \hbar} 
\int d\phi dr^2~ \delta\left[ \frac{\omega}{2}(r^2 - R^2)\right]
e^{-[r^2+2r\sqrt{R^2+q^2} \cos \phi]/ \hbar} \\ \nonumber
&= \frac{C}{\omega} ~e^{-(2R^2 + Q^2)/ \hbar} 
\int^{2\pi}_0 d\phi~ e^{-R\sqrt{R^2+Q^2} \cos \phi / \hbar} \\ \nonumber
&= \frac{2\pi C}{\omega} ~e^{-(2R^2 + Q^2)/ \hbar}~ I_0\left(\frac{2R~\sqrt{R^2+Q^2}}{\hbar}\right),
\label{posav5}
\end{eqnarray}
where $I_0$ is the modified Bessel function of zero order.

Likewise, the exact expression for the interference term, in the case of a circular shell,
is rewritten from \eref{negav} as
\begin{eqnarray}
\label{negav3}
\langle {\oprho_\pi}\rangle^- &= \frac{C}{2} \Re 
\int dpdq~ \delta\left[ \frac{\omega}{2}(p^2 + q^2 - R^2) \right]
e^{-[(p+R)^2 + q^2+2iQ(p+R)]/ \hbar} \\ \nonumber
&= \frac{C}{2} \Re  ~e^{-(R^2 + 2i Q R)/ \hbar}
\int d\phi dr^2~ \delta\left[ \frac{\omega}{2}(r^2 - R^2)\right]
e^{-[r^2+2r(R+iQ) \cos \phi]/ \hbar} \\ \nonumber
&= \frac{C}{\omega} \Re  ~e^{-2(R^2 + i Q R)/ \hbar}
\int^{2\pi}_0 d\phi~ e^{-2(R^2+iQR) \cos \phi / \hbar} \\ \nonumber
&= \frac{2\pi C}{\omega} \Re ~e^{-2(R^2 + iQR)/ \hbar}~ I_0\left(\frac{2(R^2+iQR)}{\hbar}\right).
\end{eqnarray}

In the semiclassical limit, $R^2/\hbar \rightarrow \infty$, so that one may employ the approximation 
for large modulus of the modified Bessel function \cite{Abramowitz}:
\be
I_0 (z) \approx\frac{e^z}{\sqrt{2\pi z}}.
\label{aproxbessel}
\ee
Thus, recalling the definition \eref{intersection} of the distance of the coherent state
to the circular shell, one has
\be
\langle {\oprho_\pi}\rangle^+ \approx \frac{C}{\omega}\frac{\sqrt{\pi\hbar}}{\sqrt{R^2+RP}}~
e^{-P^2/\hbar},
\ee
which, using the further approximation in \eref{intersection}, becomes identical to
\eref{posav2}. 

On the other hand, inserting (\ref{aproxbessel}) in (A.3) leads to the same
approximation for the negative interference term $\langle {\oprho_\pi}\rangle^-$ in (\ref{negav2}).
It follows that the exponential factor by which the negative average surfaces
the positive part in the semiclassical limit for a circular energy shell 
is just the same as was obtained by locally substituting the circle for its
tangents at three points.

It should be pointed out that the circular shell has also itself been obtained as a local
approximation for an arbitrary curved energy shell and it in no way implies that one is dealing
with a global harmonic oscillator. The negativity witness \eref{Wigcat} 
constructed in section 2 is purely local, because of the possibility of choosing the
separation of the coherent states to be semiclassically small, though considerably
greater than the effective radius of each coherent state. In this way, the rederivation
carried out in this appendix just confirms that curvature corrections do not affect
the results calculated using tangents.

\ack{It is a pleasure to thank Fabricio Toscano, Raul Vallejos, Gert L. Ingold and Rodolfo Jalabert  for interesting discussions. Partial financial support from the National Institute for Science and technology-Quantum Information, FAPERJ and CNPq is gratefully acknowledged.}

\end{document}